**Contrast mechanisms in secondary electron e-beam induced current (SEEBIC) imaging**

*Ondrej Dyck[1], Jacob L. Swett[2], Charalambos Evangeli[3], Andrew R. Lupini[1], Jan Mol[4], Stephen Jesse[1]*

[1] *Center for Nanophase Materials Sciences, Oak Ridge National Laboratory, Oak Ridge, TN*

[2] *Biodesign Institute, Arizona State University, Tempe, Arizona 87287*

[3] *Department of Materials, University of Oxford, Oxford OX1 3PH, UK*

[4] *School of Physics and Astronomy, Queen Mary University of London, London E1 4NS, UK*

Keywords: Secondary electron e-beam induced current, scanning transmission electron microscopy, graphene nanodevice, resistive contrast imaging, conductivity, SEEBIC, contrast, substrate effects.

## Abstract

Over the last few years, a new mode for imaging in the scanning transmission electron microscope (STEM) has gained attention as it permits the direct visualization of sample conductivity and electrical connectivity. When the electron beam (e-beam) is focused on the sample in the STEM, secondary electrons (SEs) are generated. If the sample is conductive and electrically connected to an amplifier, the SE current can be measured as a function of the e-beam position. This scenario is similar to the better-known scanning electron microscopy (SEM)-based technique, electron beam induced current (EBIC) imaging except the signal in STEM is generated by the emission of SEs, hence the name SEEBIC, and in this case the current flows in the opposite direction. Here, we provide a brief review of recent work in this area, examine the various contrast generation mechanisms associated with SEEBIC, and illustrate its use for the characterization of graphene nanoribbon devices.

# Introduction

In a scanning transmission electron microscope (STEM), one of the standard imaging modes is now high/medium angle annular dark field (H/MAADF), where the image contrast represents the number of primary electrons scattered onto the annular detector. This technique is popular because the image directly reflects the atomic structure, with intensity representing the type and number of atoms (Pennycook, 1989). While truly quantitative interpretation of this imaging mode can be subtle (Lebeau & Stemmer, 2008; LeBeau et al., 2010) it is simpler than for bright field imaging, where detailed interpretation relies upon extensive comparison to simulations (Jia et al., 2014, 2003). Techniques such as ptychography (Rodenburg, 2008)) or holography (Lichte, 1986) rely on interference patterns and computational reconstruction becomes essential. Complementary to these imaging techniques, electron energy loss spectroscopy (EELS) examines the energy distribution of the primary electrons after they have interacted with the sample (Egerton, 2011). In each of these cases, the carriers of sample information are the primary electrons.

Secondary electrons (SEs) are also emitted from the sample during interactions with the electron beam (e-beam) and contain information about the sample. SEs form the main imaging mode of a scanning electron microscope (SEM) (Goldstein et al., 2017). SE detectors have also been incorporated within a STEM. The information this imaging mode conveys is the SE yield as measured by the detector. It is worth noting that because the detector does not capture all the emitted SEs, the position of the detector becomes a major factor that influences image contrast. Certain areas can appear darker for the same SE yield if the area is shadowed due to the sample geometry and in this way, topographical information is conveyed.

Here, we will demonstrate a related imaging technique that approaches the detection of SEs, albeit with a different strategy. Rather than directly collecting SEs to image the sample, their emission is inferred by measuring the current flowing through the sample. Electron beam induced current (EBIC) imaging is an established technique that examines the current flowing through a sample induced by interactions with the e-beam, e.g., the generation of electron-hole pairs in a p-n junction. The EBIC signal examined is due to SE emission; thus, it is referred to as SEEBIC.

STEM SEEBIC is a relatively new approach that allows probing various aspects of a material such as topology, work function, electrical connectivity, and SE emission, all with the spatial resolution afforded by a modern STEM (Hubbard et al., 2020). Given that the SEEBIC technique is relatively new and many researchers may be unfamiliar with it, we have opted to compile an extended background of its development and progress. Following this short review, we will detail recent experimental observations regarding sources of contrast generation in SEEBIC.

**Review**

SEEBIC imaging is a combination of SE imaging and EBIC imaging; each method is discussed on its own merits in sections titled "Secondary Electron STEM Imaging" and "Electron Beam Induced Current (EBIC) Imaging and EBIC Taxonomy". We also discuss how these imaging modes are combined into the SEEBIC imaging mode in the section titled "Secondary Electron - Electron Beam Induced Current (SEEBIC) Imaging". Finally, recent literature is summarized in the section titled "Previous Work".

*Secondary Electron STEM Imaging*

SE imaging has been available on STEMs for several decades (Allen, 1982; Liu & Cowley, 1991; Bleloch et al., 1989; Imeson et al., 1985) was an option on JEOL STEMs in the 1980s and 1990s (Mitchell & Casillas, 2016) resolution SE imaging was first reported in 2009 (Zhu et al., 2009). This was followed by a comprehensive report in 2011 (Inada et al., 2011) and the explanatory theory was published as recently as 2013 (Brown et al., 2013). The advantage offered by SE STEM imaging, which is not achievable with conventional SEM imaging, is the resolution afforded by using accelerating voltages up to 300kV, whereas typical SEMs operate at accelerating voltages of 30-40kV (Dusevich et al., 2010). The higher accelerating voltages available in STEM enable reported SE resolutions less than 1Å (Zhang, 2011). While initial reports have attracted much attention due to their novelty and impressive achievements, relatively few publications on the topic since then suggests that this imaging mode is not heavily utilized. While merely speculative, one possible cause is a lack of required hardware on most currently installed microscopes. Another possibility is that the introduction of an SE detector may create unwanted electron-optical disturbances for the primary imaging mode. It may also be that this imaging mode simply offers little additional benefit beyond the characterization that can already be done in a typical STEM. Another limitation that is present in SE STEM imaging is that if the sample does not have a continuous electrical connection to an adequate reservoir of electrons, imaging will not be possible after some length of time, which is reported to be as short as 80 s of continuous scanning (Mitchell & Casillas, 2016). Despite this constraint, SE imaging in a STEM offers access to depth and topography information that is hard to access by either phase contrast or ADF imaging and can also provide insights into surface crystallographic information that can complement other characterization modalities (Mitchell & Casillas, 2016). It should be noted that the initial atomic resolution results were achieved with high atomic number (Z) uranium atoms from $UO_2$ (Zhu et al., 2009). Subsequent reports by the same group extended these studies to Au nanoparticles, YBCO lamella, $SrTiO_3$, and lower-Z Si dumbbells; however, there has not yet been any reports of single-atom, low-Z SE imaging; only lattice imaging (Inada et al., 2011). While such impressive results should not be downplayed, the question remains open as to whether it is possible to achieve atomic resolution SE imaging with low-Z elements, such as C, although the topic was the subject of a theory manuscript (Brown et al., 2013). Other notable work was conducted by Suenaga and colleagues to distinguish single layer graphene from single layer h-BN in a heterostructure stack with SE imaging, representing a different achievement for SE STEM imaging (Cretu et al., 2015)

*Electron Beam Induced Current (EBIC) Imaging and EBIC Taxonomy*

Electron beam induced current (EBIC) imaging is a technique that has been used widely in semiconductor fabrication (particularly for failure analysis), but has achieved less interest from the broader research community as evidenced by the number of publications. EBIC has been employed as a characterization method for over fifty years (Everhart et al., 1964) confusion over terms and mechanisms remain. Broadly speaking, EBIC techniques are used to spatially measure currents occurring within a sample caused by interactions with a focused primary beam. The most well-know EBIC technique, shown in Figure 1 (a), uses an e-beam to measure electron-hole (e-h) pair recombination lengths and can map the current depletion region in a semiconductor p-n junction. In this mode, the current measured is greater than that of the primary e-beam. The other technique, shown in Figure 1 (b), uses EBIC to probe electrical shorts in embedded traces, where the absorbed electron current from the e-beam can be used as a diagnostic tool for probing disconnects in a non-destructive manner. In this case, since the current measured is simply the absorbed e-beam current, we refer to this as electron beam absorbed current (EBAC). The current measured in this case is approximately equal to that of the primary e-beam. The third case, shown in Figure 1 (c), is of primary interest in this manuscript. Here, the measured current arises from holes generated by the emission of SEs, in what we will refer to as secondary electron EBIC (SEEBIC), following the naming conventions proposed by Hubbard et al. (Hubbard et al., 2020). The current measured in SEEBIC is generally much less than the primary e-beam current since it relies on the SE yield of the specimen.

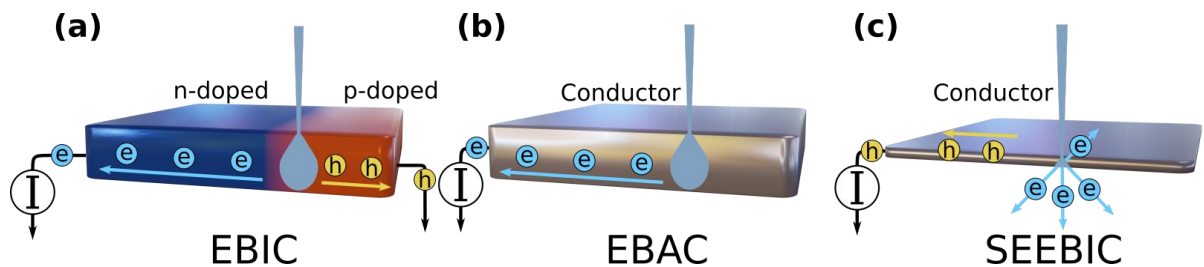

**Figure 1** Schematics of the most common EBIC concepts. (a) An incident e-beam excites holes and carriers in the depletion region of a p-n junction, resulting in a current through the transimpedance amplifier. (b) An incident e-beam is absorbed by a conductive specimen resulting in a current through the amplifier. (c) An incident e-beam generates SE emission from a thin specimen resulting in a hole current through the amplifier.

By adopting these different naming conventions, we aim to clearly delineate the main contrast formation mechanism—whether e-h pairs are generated, electrons are absorbed from the e-beam, or current arises from SE emission. Combinations of these effects cannot, in general, be neglected; however, the naming convention captures the main contrast formation mechanisms. There is also the possibility for currents to arise elsewhere and contribute to or be attributed to an e-beam induced process, e.g., when bias voltages are applied.

It should be noted that each of these mechanisms is dependent upon beam/specimen interactions, electrical connections within the sample, and electrical connections to the sample. For example, EBAC is only possible when the e-beam—or some measurable fraction of the e-beam—can be stopped within the conductive portion of the specimen. This is typically not possible during high accelerating voltage STEM imaging where almost all the electrons transmit through the sample. Likewise, SEEBIC requires the ejection of electrons from the surface, which only occurs within several nanometers of the surface. SEEBIC can attain higher imaging resolution because of the limited interaction volume, but this is at the expense of less signal current.

While a qualitative description of the mechanisms is essential to understand the observed phenomena, it is also useful to place the movement of charge on a more quantitative footing. Here, we can build on prior work and provide an equation to account for the total current in and through the system (Hubbard et al., 2018).

$$I_P + I_{IN} = I_T + I_{BS} + I_{SE} + I_A + \frac{V_S}{R_S} + I_{OUT}$$

$I_P$ is the primary incident e-beam current, $I_{IN}$ is any current injected into the system that is not attributable to the e-beam and $I_T$ is the transmitted e-beam current through the sample, where $I_T < I_P$. $I_{BS}$ represents the backscattered electron current, which, for thin samples, is effectively zero. $I_{SE}$ is the SE current leaving the sample and is the current of primary interest and $I_A$ is the absorbed current, which primarily occurs in thick samples with low primary e-beam energies. $V_S/R_S$ represents the current flowing into the sample given a non-negligible resistance $R_S$ to ground. $I_{OUT}$ is the flow of current out of the sample, balancing $I_{IN}$.

*Secondary Electron - Electron Beam Induced Current (SEEBIC) Imaging*

As mentioned in the introduction, STEM SEEBIC allows probing material properties including topology, work function, electrical connectivity, and SE emission (Hubbard et al., 2020). Because it is performed in a STEM, the tightly confined e-beam confers high spatial resolution. Crucially, SEEBIC differs from ADF imaging in that image contrast is generated from interactions with electrons in a sample as opposed to the nuclei. Most electrons are correlated strongly with the atomic nuclei, but delocalization is possible, and this can sometimes confound interpretation of SEEBIC images compared to interpreting ADF images. While in situ electrical biasing techniques also permit electrical characterization in the microscope, SEEBIC is distinguished in its ability to probe disconnected circuits.

SEEBIC is similar to SE STEM imaging in that an image is formed by scanning a focused e-beam over a sample and generating SEs; however, it differs in two key ways. First, while SEs are generated any time the beam interacts with the specimen, SEEBIC permits selective examination of SEs emitted from a region electrically connected to the amplifier. This constraint is advantageous for visualizing electrical connectivity. Second, SE STEM uses a detector that only captures a portion of the emitted SEs, sampling a small area, whereas SEEBIC typically captures a signal from all the SEs emitted, providing a detection efficiency of unity (assuming that the entire area under examination is electrically connected to the amplifier). This makes SEEBIC a more straightforward strategy for examining total SE emission and the SE yield. The requirement for a conductive path for SEEBIC does limit the types of samples that can be studied; however, as long as transport can occur, including through leakage current and other transport mechanisms, SEEBIC can be performed.

There are subtle factors that influence STEM SEEBIC contrast. These factors also complicate SE imaging in both STEMs and SEMs, as will be discussed below, but these aspects are more prominent in SEEBIC. Most of these effects arise from the interaction of the conductive SE-generating region with the (sometimes) higher SE-generating dielectrics. Mechanisms such as electron hopping and tunneling permit the exchange of electrons over "short" distances, which are well within the regimes probed by modern aberration corrected STEMs.

*Previous Work*

The most comprehensive and descriptive previous work is from 2018 (Hubbard et al., 2018) clearly describes what is now recognized as SEEBIC and presents a clear case that distinguishes this technique from more traditional forms of EBIC. Another notable report was published in 2020 (Hubbard et al., 2020)builds on work from the previous report and demonstrates the technique of STEM-EBIC, including SEEBIC, and shows that by using clever device design and focused ion beam (FIB) sample preparation, regions of interest in devices can be selectively extracted and analyzed with SEEBIC. A significant advance in 2019 showed that 2Å resolution was achieved with SEEBIC (Mecklenburg, Hubbard, Lodico, et al., 2019). In this work, lattice-resolution images were acquired of Au nanoparticles on silicon nitride membranes. A recent publication in 2021 showed that the SEEBIC technique was sensitive enough to distinguish single layers of suspended graphene despite its very low SE yield (Dyck, Swett, Lupini, et al., 2021). Utilization of the SEEBIC technique for characterizing graphene-based nanoelectronic devices was also presented in 2021, where the authors noted SEEBIC-based detection of conductance switching in graphene nanoconstrictions (Dyck, Swett, Evangeli, et al., 2021).

Another publication is notable for implementing resistive contrast imaging (RCI) to study a component from an actual device, in this case a multilayer ceramic capacitor (MLCC) (Hubbard et al., 2019). The authors visualized resistive grain boundaries in a sintered polycrystalline $BaTiO_3$ dielectric layer from a capacitor, demonstrating heterogeneity in the material, which has implications for device performance such as reliability and the dielectric constant. The work shows how conductivity variations within a specimen, along with an understanding of the charge production mechanism, can facilitate mapping the resistivity of a thin specimen with high spatial resolution that not only improved the resolution previously demonstrated with RCI using EBAC (Russell & Leach, 1995), but also demonstrated how SEEBIC enables RCI in STEM whereas EBAC is not possible due to the low absorption probability.

In another experiment, STEM EBIC was used to study perovskite solar cell components *in situ* using protective electrodes for air-sensitive specimen components (Duchamp et al., 2020). This report focuses primarily on the induced current generation mechanisms. The authors have a narrower definition of EBIC, defined as charge carrier generation in an electrically contacted semiconductor device (traditional EBIC), but acknowledge and describe contributions to the signal from SEs. The relative contributions of the SE signal are separated from the top and bottom of the sample, which they were able to independently probe and show that the EBIC signal in the STEM matches surface sensitive SE imaging in an SEM, thus verifying the contribution of SEs to the total EBIC signal.

These and other related STEM EBIC studies (White et al., 2015; Sparrow & Valdrèg, 1977; Brown & Humphreys, 1996; Fathy et al., 1980) provide a sense for the current status of the field. The most significant developments of importance to the present work have been briefly described, providing the foundational understanding that is built upon here.

## Results

Having described the current state of developments and research in the field, a number of experiments and observations using SEEBIC for the examination of graphene nanoelectronic devices will be presented. These are divided into the following sections: in "Materials and Methods" sample fabrication, preparation, and acquisition details are described; in "Imaging Supported Graphene" we show that a single layer of graphene is easily detected when supported on a $SiN_x$ substrate; in "Substrate Contributions" the role of the substrate in generating the observed SEEBIC contrast is discussed; in "*In Situ* Device Diagnostics" we show how the SEEBIC technique can be leveraged to reveal electrical connectivity; in "Influence of Voltage Bias on Image Contrast" a series of electrical biasing experiments are presented and the SEEBIC contrast is examined; in "Resistive Contrast Imaging" evidence for RCI in supported graphene is presented.

**Materials and Methods**

A conductive path to the region of interest on the sample is necessary to acquire SEEBIC images in STEM. To facilitate this, a custom sample platform was designed and fabricated featuring lithographically patterned electrodes that interface with the electrical contacts on the Protochips$^{TM}$ TEM holders. An extensive description of the fabrication process is presented in the supplemental information of a previous publication (Dyck, Swett, Evangeli, et al., 2021). Briefly the platform consists of a 300 μm thick Si base, 1000 nm of thermal oxide to reduce capacitive coupling between the electrodes and the Si base, and 20 nm of low stress $SiN_x$. Electrodes were patterned on top using two lithography steps: fine features were patterned using e-beam lithography and metalized using e-beam evaporation (Cr 5 nm/Au 35 nm) and larger features were patterned using photolithography and metalized using e-beam evaporation (Cr 5 nm/Au 95 nm). A back side etch using KOH at 80 °C was used to form the electron transparent window located beneath the fine electrodes. For samples with apertures in the window, a Ga+ FIB was used to mill away the $SiN_X$ membrane. Graphene was transferred to the full wafer by Graphenea (San Sebastián, Spain) and patterned using e-beam lithography.

Imaging was performed using a Nion UltraSTEM 200 operated at 60, 100, and 200 kV accelerating voltages (as indicated in the text). Beam currents are also listed individually. The nominal convergence angle was 30 mrad. A custom-built SEEBIC acquisition platform was used to record the SEEBIC signal, consisting of a break-out box to make and break electrical connections to the sample and a Femto DLPCA-200 transimpedance amplifier (TIA). All SEEBIC images were acquired with a gain of $10^{11}$ V/A and a bandwidth of 1 kHz. The TIA output was connected to an unused analog input of a Gatan DigiScan to facilitate parallel acquisition of SEEBIC and high angle annular dark field (HAADF) signals. Since digitization of the signal by the DigiScan system is opaque, the numerical value of the image intensity is not directly interpretable as a measured current. To obtain quantitative current measurements the TIA output was disconnected from the DigiScan system and connected to a custom-built data acquisition platform (Li et al., 2018; Sang et al., 2016, 2017) to scan the e-beam and record the various signal outputs. Analysis of the images acquired with this custom-built system are shown with intensities of pA, whereas analysis of the images acquired with the DigiScan system are shown with intensities of arbitrary units (a.u.).

The SEEBIC signal exhibited a significant interference component, which was synced with the electrical mains manifesting as vertical stripes. This signal was removed from all images presented here in a procedure fully described in the supplemental information of a previous publication (Dyck, Swett, Lupini, et al., 2021; Dyck, Swett, Evangeli, et al., 2021). Briefly, a script was used to sum the images vertically and obtain a moving average that represents the larger scale image features (background intensity). The background signal was removed by dividing the original signal by the background. The remaining signal represents the interference pattern to be removed. The original signal was then divided by the interference signal to obtain a corrected version of the SEEBIC image.

**Imaging Supported Graphene**

SEEBIC images are largely amenable to intuitive interpretation, especially when focused on the conductive elements; however, there are subtle aspects that should be included. Here, new interpretations of previous SEEBIC results are introduced that are more readily understood because of the 2D nature of graphene. This section will provide clarity for image interpretation by detailing the mechanisms behind image formation and contrast.

An illustrative device is shown in

Figure 2, where a SEEBIC image, (a), of a partially suspended graphene sheet was acquired through one electrode while the other, disconnected electrode, was grounded. This image can be compared to the simultaneously acquired HAADF image, (b), noting a few aspects that prominently stand out. First, the graphene is clearly visible on the substrate, which is worth noting since the graphene is only a single layer of carbon atoms while the $SiN_x$ is nominally 20 nm thick. As far as the mass thickness of the sample is concerned—the contrast formation mechanism of ADF imaging—the graphene represents a negligible component. However, when balancing charge flow, graphene supported on $SiN_x$ is a starkly different electronic system than bare $SiN_x$ and the SEEBIC contrast is primarily governed by this balancing of charge.

Second, the connected Au electrode is bright, as is common in ADF imaging, but the grounded electrode has the same intensity level as vacuum. The dark contrast of the grounded electrode can be explained by the fact that when the e-beam strikes this electrode, SEs are generated, but electrons are replenished by a direct connection to the common ground (not through the TIA), and therefore, no current is measured.

Third, the graphene has a lower SEEBIC signal in suspended regions compared with supported regions. This difference in contrast, shown clearly in

Figure 2 (c), (e), and (f), has a less intuitive explanation as it is due to the substrate contribution of the supporting $SiN_x$, which we will discuss in detail in the next section. Another subtle feature is a lack of sharp contrast between the graphene and the $SiN_x$ substrate, which is another substrate contribution to the SEEBIC signal.

Finally, a strong signal at the edge of the aperture is an additional feature present in the images that will also be explained in the next section, but can be understood as the same edge effect observed in common SEM imaging (Seiler, 1983).

To better visualize the SEEBIC signal, a 3D surface plot is included in Figure 2 (d); this highlights the lack of a defined edge at the graphene/substrate interface and shows the difference in SEEBIC signal between the Au electrode and supported graphene. To put these intensities on a quantitative footing, we analyzed the intensities as follows. The SEEBIC image was segmented into different categorical regions based on feature finding using the trainable Wekka segmentation plugin in Fiji (Schindelin et al., 2012; Arganda-Carreras et al., 2017). Histograms for each region are show in

Figure 2 (e). The mean values are compared in the bar chart shown in

Figure 2 (f), with the standard deviation shown as error bars. The mean vacuum signal was used as a zero reference and all intensities (current values) were adjusted so that the mean vacuum reading was zero. The closest distinct signals are those of the substrate and vacuum and, as anticipated, a direct overlap exists between the SEEBIC intensity for vacuum and the separately grounded contact. The mean current values obtained for each region are listed above the corresponding bars in pA.

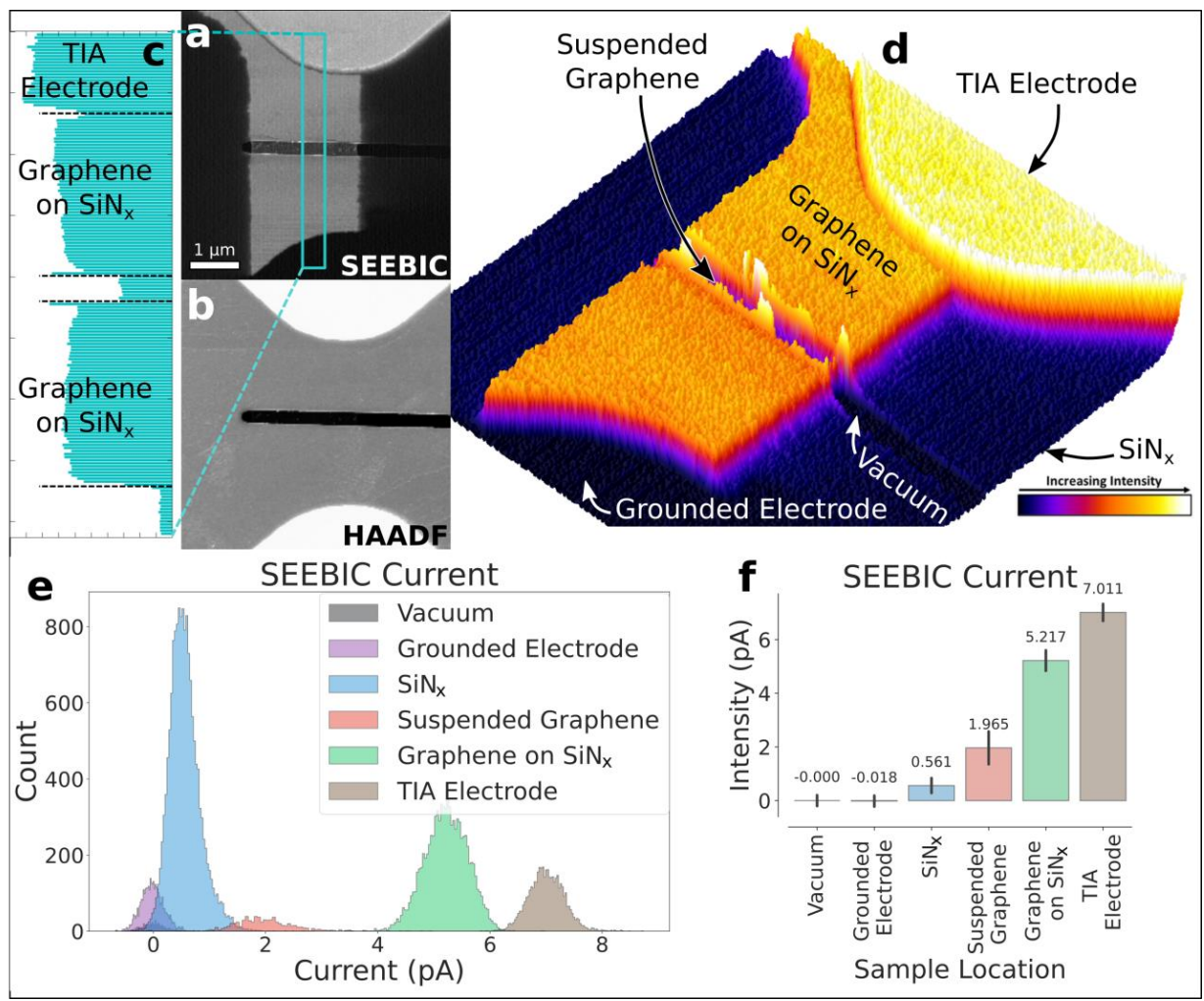

**Figure 2** A graphene ribbon device, partially suspended and severed from contact with the bottom electrode, showing several image contrast mechanisms. a) Greyscale SEEBIC image showing the grounded and contacted electrodes and b) concurrently acquired HAADF-STEM image. c) Profile across the device showing the ability of SEEBIC to distinguish graphene on $SiN_x$, suspended, and grounded contact. d) 3D surface plot of a) highlighting differences in signal. e) Histogram of SEEBIC intensities in pA from various regions. f) mean current for various regions. Image acquired at 60 kV with a 13.3 ms pixel dwell time.

**Substrate Contributions**

The relative SEEBIC image intensity decreases when graphene is suspended compared to locations where it is supported on the substrate. One might assume that dielectric materials would not contribute to SEEBIC contrast formation and can therefore be neglected from further consideration, but this is not the case. In this section, the contribution of the substrate to SEEBIC imaging is examined and the likely mechanisms are proposed. First, it is worth noting that substrate contributions to SE signals from nanomaterials have been discussed and debated in the context of SEM imaging. Substrate contributions are more substantial when the specimen being imaged is comprised of a single atom, atomically thin layer, or otherwise has a dimension on the order of <5nm, which is the approximate escape depth of SEs.

In 2010 a controversial report published an anomalously high SE emission from carbon nanotubes (Luo et al., 2010). Shortly after, a comment was published by another group suggesting that the images used to support this claim were misinterpreted and that the electrons emitted were likely coming from the substrate, with the carbon nanotubes able to replenish the substrate electrons in the vicinity, accounting for a larger SE yield from that region (Alam et al., 2011). In their conflicting study of the SE yield of multiwalled carbon nanotubes (MWCNTs), they suspended the MWCNTs and obtained a much lower SE yield. The original group published a response (Luo et al., 2011) The underlying debate concerns the role of the substrate in the SE emission process. Challenges regarding the interpretation of substrate effects are not limited to this one instance (Park et al., 2010; Park & Ahn, 2015). A common mistake is that when SEEBIC is used to probe conductivity, only ohmic conductivity is considered, whereas in reality, all forms of conduction, especially those typically considered leakage current, also play a role.

With new experimental data from STEM SEEBIC, additional insight can be provided.

Figure 2 shows a SEEBIC image of graphene suspended across an aperture in the thin $SiN_x$ membrane. In the suspended regions substantially lower current is observed compared to the supported regions. This difference is clearly highlighted in the line profile in Figure 2(c) and the quantitative comparisons shown in Figure 2 (e) and (f).

Conduction should be examined through the lens of the graphene/$SiN_x$/e-beam system. Silicon nitride is a dielectric with a resistivity in the range of $10^{10}$-$10^{17}$ Ω-cm (Joshi et al., 2000). The wide range can be attributed to differences in stoichiometry, quality, and stress, which can drastically vary the resistivity likely through Poole-Frenkel (PF) transport (Blázquez et al., 2014) and can result in orders of magnitude differences in resistivity for relatively small changes in stoichiometry (Habermehl & Carmignani, 2002). Additionally, silicon nitride has a bandgap of ~5 eV; however, this can decrease to ~2 eV with increasing Si content. Here, experiments used Si-rich $SiN_x$; thus, it is reasonable to assume that the band gap was between 3-4 eV (Tan et al., 2018). The conduction of current in such an insulator, with fields below the bandgap, is expected to be very small; however, this should not be interpreted as completely prohibiting the conduction of current, particularly at high electric fields (Blázquez et al., 2014) such as those generated by extreme charging of a material due to the emission of SEs. The local electric field induced by the e-beam is not known; however, given the number of SEs emitted from such a small region of material, it is reasonable to assume that it is sufficiently high (Nan Jiang, 2016) to permit PF conduction. When a high electric field is applied, several conduction mechanisms become possible, such as PF conduction (sometimes called PF emission), hopping conduction, and tunneling over distances of a few nanometers. Studies of metal-insulator-metal (MIM)/silicon-rich $SiN_x$ show that when a 10 V bias is applied across a 20 nm thick membrane, a resistivity of approximately $10^3$ Ω-cm is measured (Graf et al., 2019).

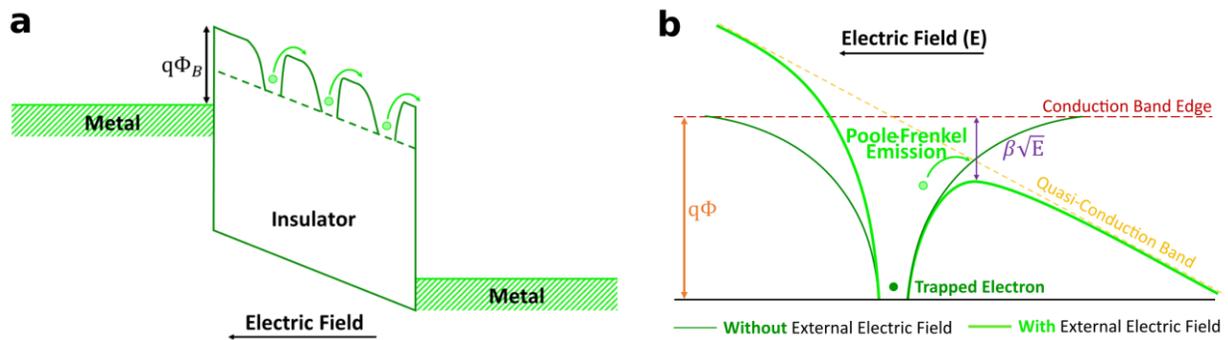

**Figure 3** a) Common band diagram of a MIM stack demonstrating PF emission, b) detailed band diagram demonstrating PF emission, where qΦ is the ionization potential and β√E is the amount by which the trap barrier height is lowered by the applied electric field (E) and β is the PF field lowering coefficient.

The most plausible candidate for conduction in the $SiN_x$ membranes examined here is PF conduction, although it should be noted this has been called into question in recent years (Schroeder, 2015) and has been suggested that Fowler-Nordheim tunneling can dominate at very high fields for some stoichiometries (Paloura et al., 1991). Under appropriate conditions, PF conduction can lead to substantial transport (Habermehl & Carmignani, 2002) and was studied in detail as far back as 1967 (Sze, 1967). With PF conduction, the substrate can be viewed as a system containing shallow trap states, as illustrated in Figure 3, where the band structure depletes with increasing voltage and the trap level rises with increasing temperature, both of which serve to facilitate conduction. These trap states are believed to arise from dangling Si bonds, which are more prevalent in Si-rich materials (Dupont et al., 1997; Robertson & Powell, 1984), although impurities from the deposition process may also contribute (Bailey & Kapoor, 1982). It is well-documented in the literature that the field dependent leakage current can vary up to seven orders of magnitude with changes in $SiN_x$ stoichiometry (Habermehl, 1998; Blázquez et al., 2014; Graf et al., 2019). The leakage current—which is both field and temperature dependent—increases with increasing silicon content of the silicon nitride, due to changes in the local atomic strain and trap density (Habermehl, 1998). Specifically, the reduced bond strain in high Si content silicon nitrides (i.e., $SiN_x$, where x<1.33) leads to a reduction in the charge trap barrier height. This relationship can be quantified by the following equation for the PF emission current density, J, as a function of electric field, E, and temperature, T.

$$J(E,T) = C_1 E e^{-q(\Phi_B - \sqrt{qE/\pi\varepsilon})/kT}$$

where the pre-exponential factor, C, is a fitted parameter and is determined by the charge trap density and the carrier mobility. $\Phi_B$ is the charge trap barrier height, ε is the high frequency dielectric constant, q is the charge quantum, and k is Boltzmann's constant. With this understanding and the knowledge that Si-rich $SiN_x$ was used in these experiments, we can now examine the image formation through the lens of a field-induced high-leakage current substrate able to contribute electrons over short distances in a graphene/$SiN_x$ system. Figure 4 shows a schematic of the e-beam interacting with both the graphene and the $SiN_x$ substrate, leading to a higher signal. This arises from the combined graphene and substrate SE emission, since single layer graphene is not sufficiently thick to stop all SEs emitted from the substrate. Additionally, the substrate will emit SEs from the side opposite the graphene, which can also be replenished by the grounded graphene, further increasing the current. The holes in the substrate—or alternatively electrons in the graphene—will readily migrate the short distance (<20 nm) to the graphene and contribute to the signal, given the high bias induced by the e-beam stripping charge locally from the specimen. This effect will be further enhanced if heating is occurring simultaneously. The exact depth from which SEs can be emitted from $SiN_x$ is unclear; however, it is likely that the probability of emission is substantially reduced inside the membrane, given reported literature on SE emission depths (Voreades, 1976). Thus, we can anticipate that SEs are predominantly emitted from the surfaces, including the edge of the aperture. This is consistent with the prominent edge effect observed in the SEEBIC signal in

Figure 2(b) and (c). A slight oversimplification in our schematic in Figure 4(c) and (d) is the sharp cut-off we have drawn for the SE emission regions. In reality, this is an exponential distribution that decays as a function of depth with no sharp cut-off. The probability for emission as a function of depth, z, is given by (Lin & Joy, 2005):

$$p(z) = ke^{-z/\lambda}$$

where $k = 0.5$ for symmetrically scattered SEs.

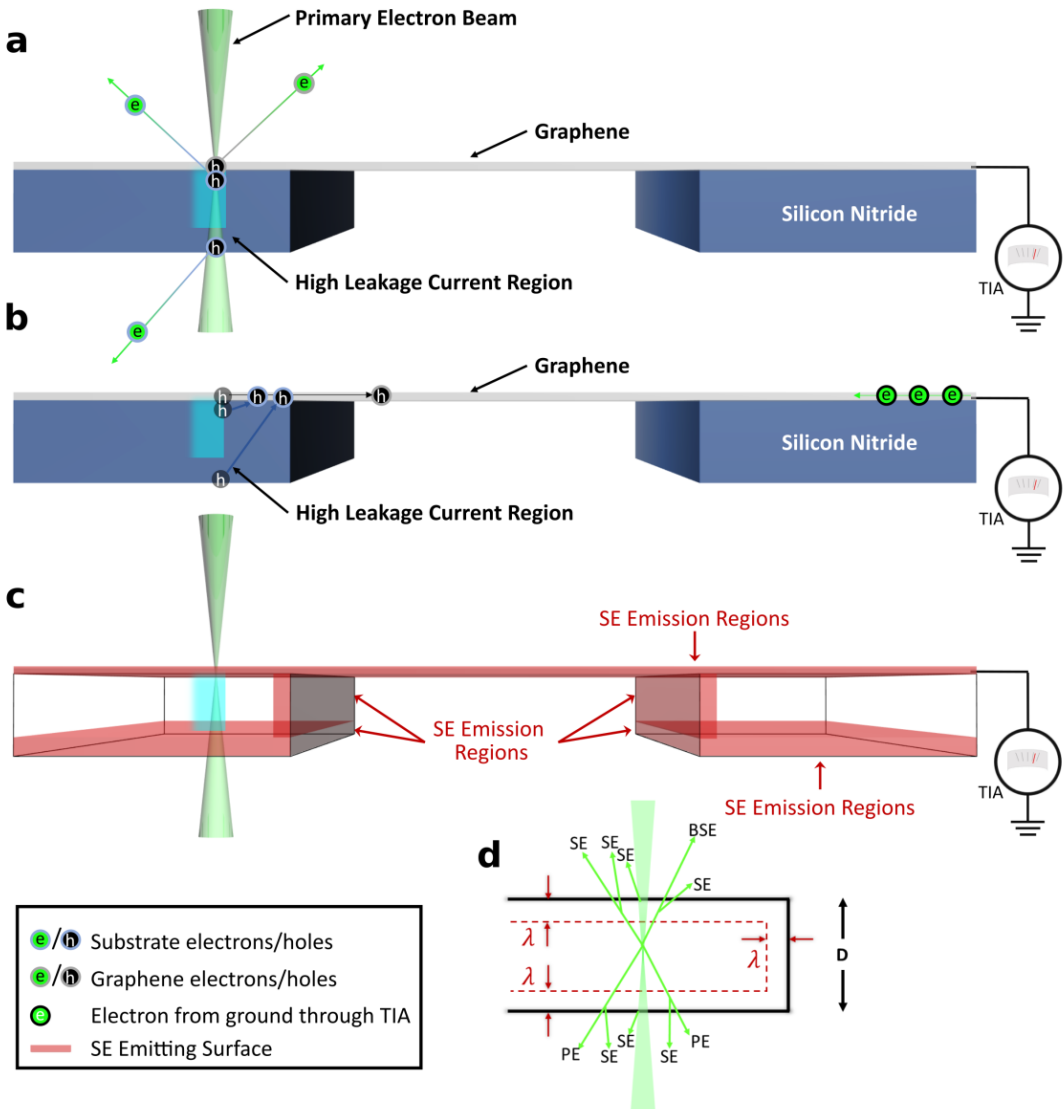

**Figure** 4 Schematic showing how an insulating substrate with a conductive layer on one side, such as graphene, can contribute to the SEEBIC signal through localized charge transport. a) Primary beam emitting SEs from both the graphene and substrate. This generates holes that are replenished by electrons from ground through the TIA. b) Migration of charge resulting in a measured current through the TIA. c) Schematic showing SE emitting regions of the device in red, which occur primarily at the outer 3-5 nm. d) Schematic showing SE emission as a function of both primary electrons and backscattered electrons, from within the distance, λ.

Figure 5 shows schematic representations of the system at various stages during data acquisition that better illustrate how these effects combine to create the relative contrast in the recorded images and details changes in signal strength expected from the various regions of the device. Figure 5(a) and (b) are explained directly by Figure 4; however, Figure 5(c) and (d) merit further discussion. The SEEBIC signal from supported graphene is a combination of the signal from the top side of the graphene and the substrate. When the graphene is suspended, SEs can be emitted from the top and bottom surfaces; thus, the difference in the signal from suspended and supported graphene is due to the absence of the substrate contribution but the addition of emission from the bottom of the graphene. Finally, when the e-beam approaches the edge of the $SiN_x$ aperture, Figure 5(d), some portion of the emitted SEs can be re-captured by a trajectory colliding with the substrate, which then migrates back to the graphene, serving to reduce the net outflow of charge and a decrease in signal.

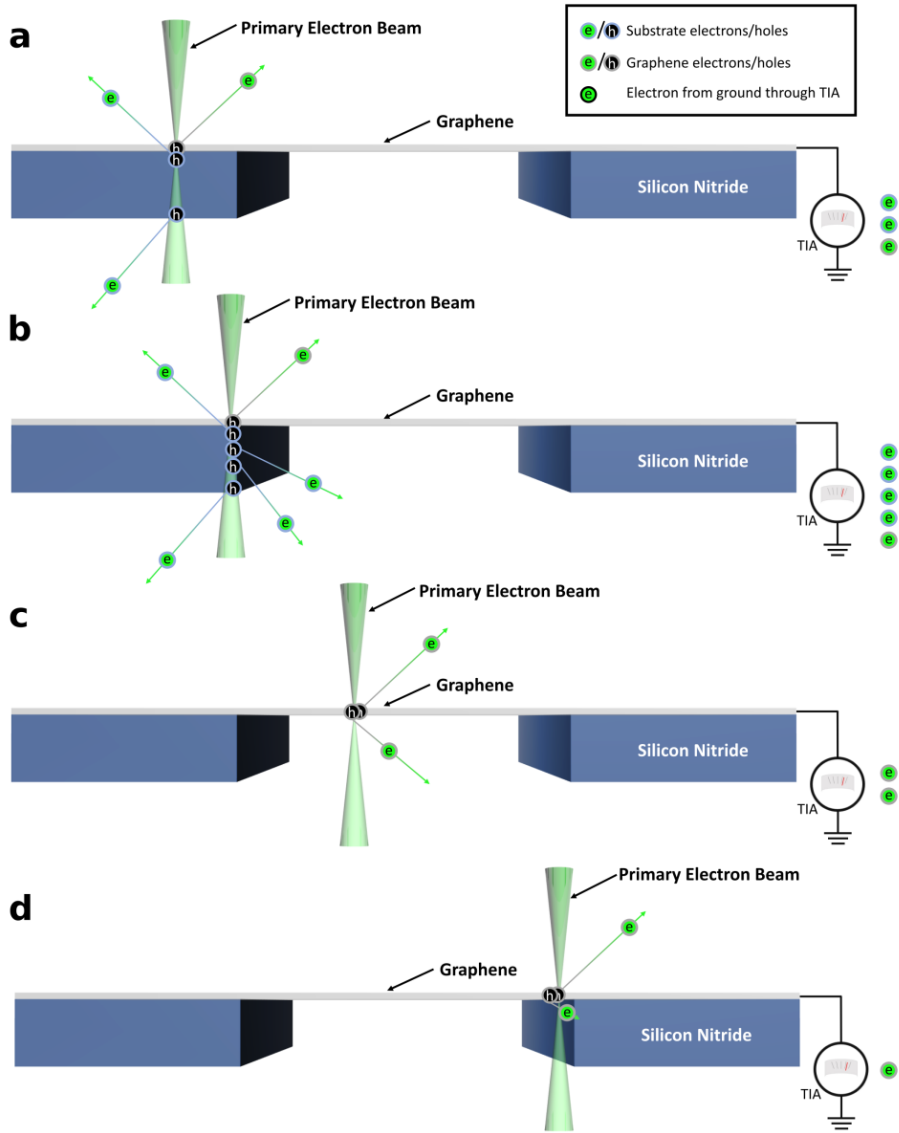

**Figure 5** Schematic showing the mechanisms behind the different intensities measured in SEEBIC images as the graphene transitions from supported to suspended. a) When the e-beam is on the graphene on $SiN_x$, signal comes from SEs emitted from both the graphene and $SiN_x$; b) at the edge of the $SiN_x$, electrons can also be emitted from the full thickness of the $SiN_x$ resulting in a substantially higher SE signal; c) when the e-beam is on the suspended graphene, only the graphene and any contamination contribute to the SEEBIC signal; d) finally, as the e-beam is on the graphene, but near the $SiN_x$, some SEs (approximately ¼) will have a trajectory that leads them to be absorbed by the substrate, resulting in decreased signal.

So far, the discussion has focused on the contribution of the substrate to the SEEBIC signal intensity where the graphene is present on the surface. We will refer to the movement of charge in this situation as vertical conduction. Since $SiN_x$ is conducting charge carriers vertically and driving the increase in SEEBIC intensity, one would expect to be able to observe similar conduction-driven effects in the lateral direction at the edges of the graphene. Indeed, these effects can be prominently observed at the graphene edges, extending 100s of nanometers from the edge in some instances. Figure 6 clearly shows this feature, which can also be observed in

Figure 2.

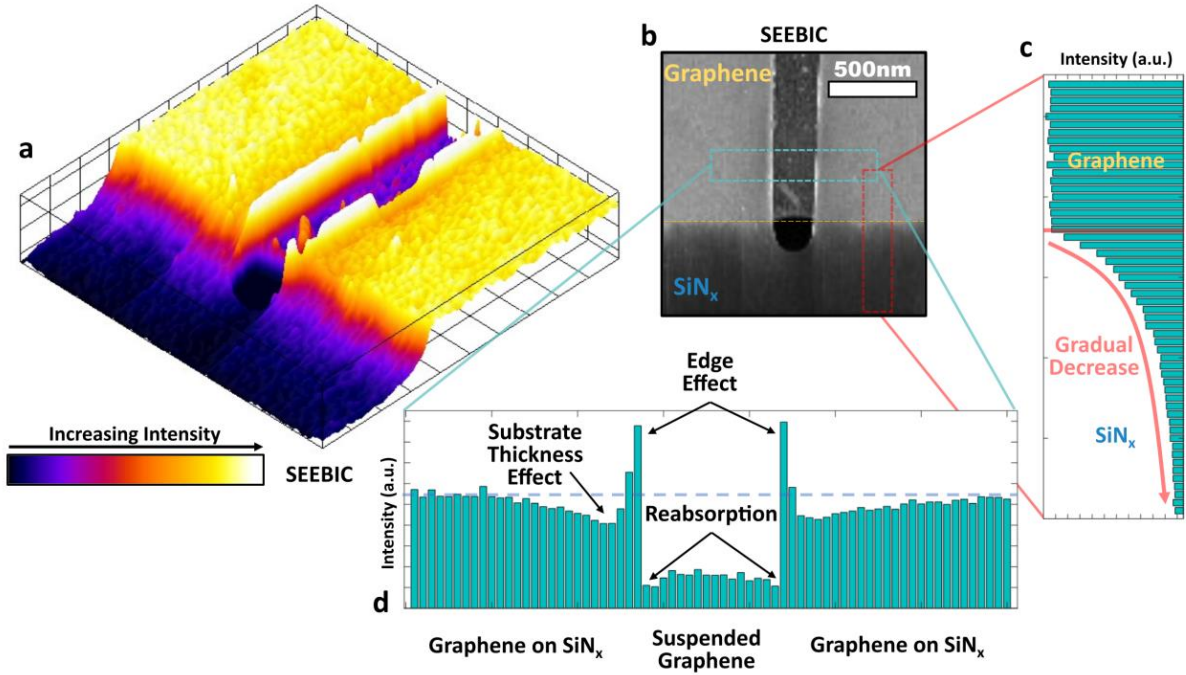

**Figure 6** Lateral conduction in the substrate. Images acquires at 60kV. a) 3D surface plot showing the increase in SEEBIC signal towards the graphene, b) greyscale SEEBIC image showing the drop off in contrast away from the graphene on the $SiN_x$, c) integrated line profile of intensity showing SEEBIC signal over 100s of nanometers, and d) integrated line profile across the aperture.

The lateral extent of the SEEBIC signal from the edge of the graphene is shown in Figure 6. The lateral conduction has been observed to extend up to 500 nm away, which is observed as a blurring of the defined edge of the graphene. The effect is not unique to graphene and is observed with any conductor, such as the metal electrode shown in Figure 7. Every SEEBIC image displayed this lateral transport in the substrate in proximity to a connected conductor. Sharp edges were only observed when there was an additional conductive region nearby with a separate ground. Integrated line profiles shown in Figure 7 demonstrate this effect by contrasting the change in SEEBIC intensity between the graphene-to-substrate transition and the graphene-to-graphene transition, where each side is separately grounded.

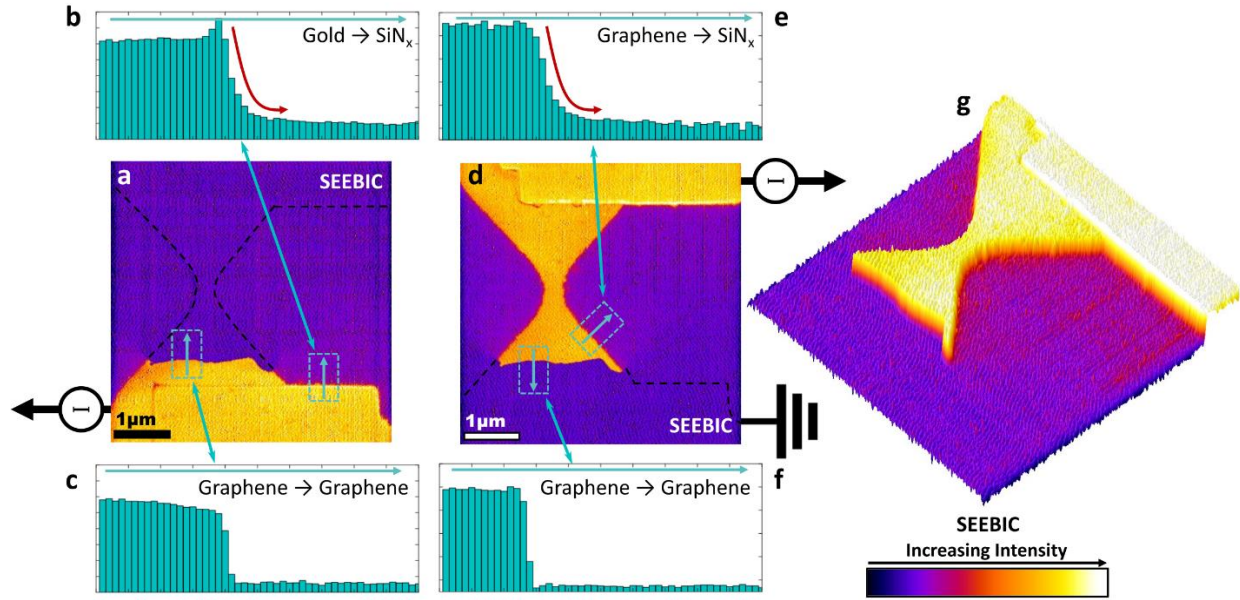

**Figure 7** Images acquired at 60kV and 57pA beam current. a) SEEBIC image of severed graphene device with electrical contact shown, b) integrated line profile of graphene edge to SiN$_x$ showing gradual transition, and c) integrated line profile of contacted graphene to grounded graphene, showing abrupt transition. d-f) same as a-c) with TIA and ground connections reversed. g) 3D surface plot of d) where the regions of substrate contributions are readily visualized.

This demonstrates that e-beam induced transport in substrates is non-negligible in SEEBIC studies. The implications of these findings also extend beyond SEEBIC to SE STEM imaging in general.

**Diagnosing Devices *In Situ* via SEBIC**

In a previous publication we showed how the SEEBIC technique could be used to probe conductivity and connectivity as a diagnostic tool in the examination of graphene nanoelectronic devices (Dyck, Swett, Evangeli, et al., 2021). In that publication it was clear that electrical continuity could be probed with high spatial resolution, which is useful in diagnosing simple device failure modes, such as for the electrically discontinuous graphene device shown in Figure 7.

In this section, we will consider more subtle behaviors where SEEBIC can also be useful.

Figure 8 shows HAADF and SEEBIC images of a partially suspended graphene device. This device showed high resistivity prior to severing the graphene through the continued application of a voltage bias and consequent joule heating (Sadeghi et al., 2015; Prins et al., 2011; S. Lau et al., 2014). Examination of the failed device with SEEBIC showed conduction in locations around the aperture in the $SiN_x$, which were not part of the initial, partially suspended, bowtie-shaped device. The conduction path around the rim of the $SiN_x$ aperture was created through temperature-induced graphitization of hydrocarbon-based contamination that accumulated at the perimeter of the aperture from the graphene transfer process. Electron energy loss spectroscopy (EELS) confirmed the presence of carbon along the $SiN_x$ aperture edges. Prior to SEEBIC characterization, these devices were assumed to have only a single conduction path in the location where the patterned graphene was placed. This result indicates that this may be inaccurate, particularly after device operation.

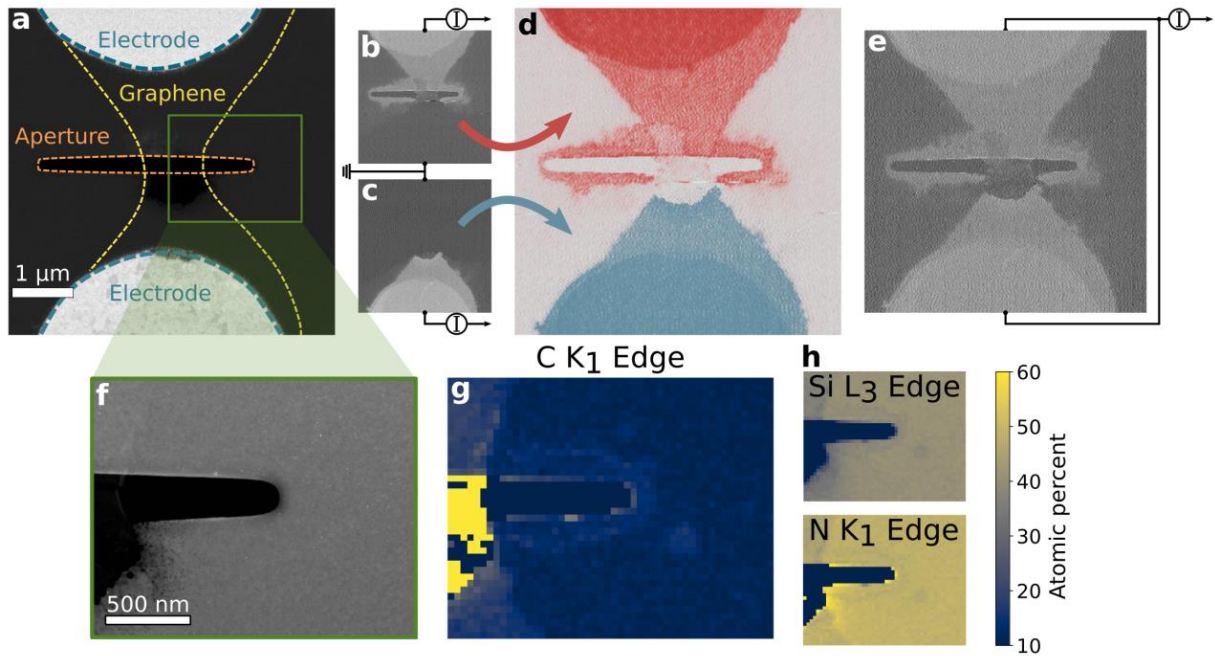

**Figure 8** SEEBIC diagnosis of device failure. Images acquired at 60 kV and 58 pA beam current. a) HAADF-STEM image of the device with various features labeled. b) and c) SEEBIC images acquired with the TIA connected to the top side of the device while the other is grounded, b), and vice versa, c). d) Artificially colored composite of b) and c). e) Both sides connected to the TIA showing that central portion of graphene device is no longer electrically connected. f) HAADF-STEM image of the EELS acquisition location. g) Relative quantification of the EELS C-$K_1$ core loss edge. Color corresponds to atomic percent relative to the Si and N EELS edges. The C track around the edge of the aperture is clearly visible. h) Corresponding quantifications of the Si-$L_{2,3}$ and N-$K_1$ EELS core loss edges. The color bar corresponds to all three EELS images.

An additional benefit of device characterization via SEEBIC is parallel M/HAADF acquisition. While ADF imaging is still the primary imaging technique and to some extent is always being acquired along with SEEBIC imaging, the parallel acquisition and subsequent overlay capability can prove particularly useful for understanding device failure. This utility arises from the complementary aspects that each technique probes and the insight gained from the combination of the two. In Figure 9, a graphene device was joule heated to the point of failure. In this instance the heating was significant enough to create an aperture and a crack in the $SiN_x$ substrate. The HAADF-STEM image provides substantial insight into the device damage, including what appears to be the outline of graphene in the $SiN_x$ substrate and a line-like feature running across the image. By correlating these images with the SEEBIC image, we can immediately see that the substrate damage can be correlated to the edge of the conductive region of the graphene, providing insight into the heating and heat dissipation at the graphene edges.

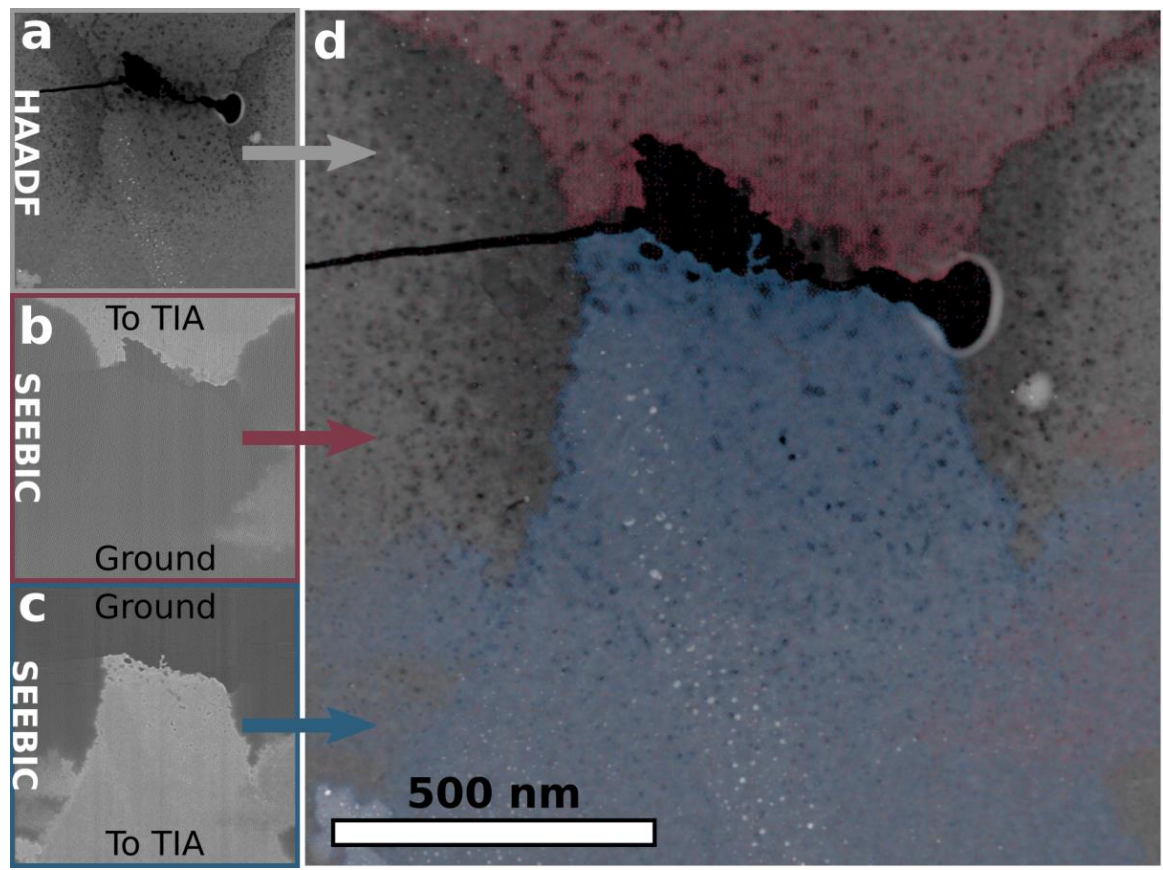

**Figure 9** Example of how simultaneous HAADF/SEEBIC imaging can help understand device failure, a) greyscale HAADF-STEM image, b) SEEBIC image of the top half of the device, c) SEEBIC image of the bottom half of the device, and d) false colored composite image showing how the conductive graphene aligns with the substrate damage, providing evidence for graphene edge-associated damage at high current densities. Images acquired at 200 kV and approximately 600 pA beam current.

In the final examples of this section, two additional sets of results are presented that point to additional avenues where SEEBIC may prove useful. The first is shown in Figure 10 where we observe an interesting difference between the SEEBIC images and HAADF-STEM images. In the two HAADF-STEM images, Figure 10(e) and (f), a bright square is observed that experienced microscopists will recognize as e-beam induced hydrocarbon deposition at the edges of a scanning frame (in this instance the bright square occurred from a long dwell-time scan during EELS acquisition). Modification of the sample during experiments is rarely desirable and such was the case in these experiments where we intended to probe conductivity *in situ* through high-resistance regions and the observed contamination could form an alternate conduction path, particularly if it was graphitized (Rummeli et al., 2018). Fortunately, SEEBIC can readily inform us if the contamination is conductive and forming a connection to the device. The contaminated region did not show up brightly in the SEEBIC images, Figure 10 (a-d), indicating that the contamination is either non-conductive or is disconnected from the device. Given that the $SiN_x$ conductivity is sufficient to produce substantial changes in the SEEBIC signal it seems most likely that the contamination is disconnected. In fact, these regions are slightly *darker* in the SEEBIC images, implying that less current is being measured on the graphene device and suggesting that substrate SEs are being replenished from this region, but on the backside of the sample (i.e., not the device side). While this possibility might have been deduced through other mechanisms, SEEBIC imaging showed clearly that it was disconnected from the device.

Another interesting feature in these SEEBIC images is the existence of a dark region in the graphene at the corner of the metal electrode (indicated by the arrows in Figure 10). The HAADF-STEM image gives no indication that this region is any different from the rest of the contact edge. A likely explanation is that there is a portion of graphene that is suspended as it stretches from the Au contact to the $SiN_x$ surface. In

Figure 2 and Figure 6, the signal from suspended graphene is lower than that from the supported graphene due to substrate contributions. Likewise, in

Figure 2 a pronounced decrease in SEEBIC intensity is observed at the border between the TIA electrode and the graphene on $SiN_x$. This is consistent with the idea that the graphene in this region is suspended above the contour formed between the electrode and the $SiN_x$ substrate. Moreover, this decrease is not consistently observed from one sample to the next or from one location to another, as shown in Figure 10, which is also to be expected if this phenomenon is a result of graphene suspended above the substrate. This data is not conclusive, but it is suggestive and provides enough impetus to explore this possibility further.

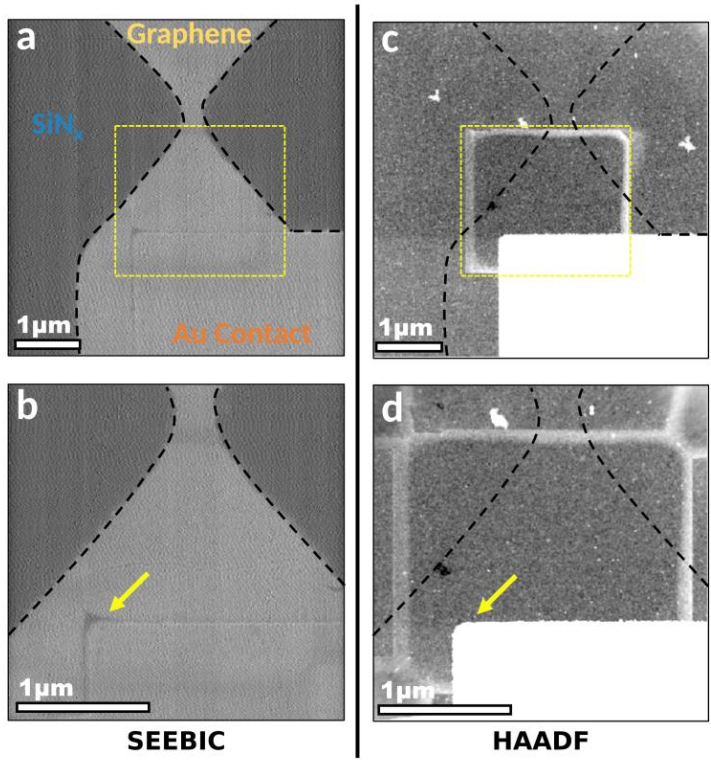

**Figure 10** Example of SEEBIC being used to confirm that contamination is not in contact with the graphene. Images acquired at 60 kV and 49 pA beam current. a) SEEBIC image of graphene device on $SiN_x$, b) higher magnification SEEBIC image. c,d) HAADF-STEM images where contamination on the backside of the device is clearly visible, which does not appear in the SEEBIC images.

**Influence of Voltage Bias on Image Contrast**

Thus far, the results presented herein were acquired with no additional bias applied to the system. The electrode disconnected from the TIA was either grounded or was floating. The two-electrode nature of the system permits further experimentation and an opportunity to validate our understanding of the physics involved in contrast formation. Here, several experiments are described where a bias was applied to one electrode (non-sensing) and the SEEBIC signal was collected from the other (sensing) electrode. These experiments provide additional insight into the physics of SEEBIC imaging, corroborate results on enhancing SEEBIC contrast (Mecklenburg, Hubbard, Chan, et al., 2019), and demonstrate a new phenomenon that we have termed secondary electron electron beam absorbed current (SEEBAC). There is a distinction between this process and the recapture illustrated in Figure 5(d) and mentioned briefly by others (Hubbard et al., 2018). Recapture leads to a reduction of the measured current, whereas SEEBAC involves a reversal in the direction of charge flow. SEEBAC is interesting in that until now, EBIC techniques required a direct connection to the TIA, but here connectivity is not required, which may enable new avenues for SEEBIC. To help clarify the differences between the techniques discussed so far, Figure 11 demonstrates how the imaging modality depends on where the e-beam is positioned and the polarity of the bias. From this we can delineate the conditions necessary for SEEBAC to occur, namely, an electric field directing emitted SEs from a negatively charged specimen to a nearby grounded specimen with a connection to the TIA (the sensing electrode), Figure 11(a). Figure 11(c) illustrates how switching the polarity results in recapture of the electrons on the negatively charged specimen, and hence no image is acquired. Finally, we observe that in the instance when the non-sensing electrode has a positive bias (e.g., attracting SEs), some recapture is prevented, leading to a slightly enhanced signal, Figure 11(d). In Figure 11(e), both SEEBIC and SEEBAC can be observed on the same specimen under the same sample conditions depending on the e-beam position.

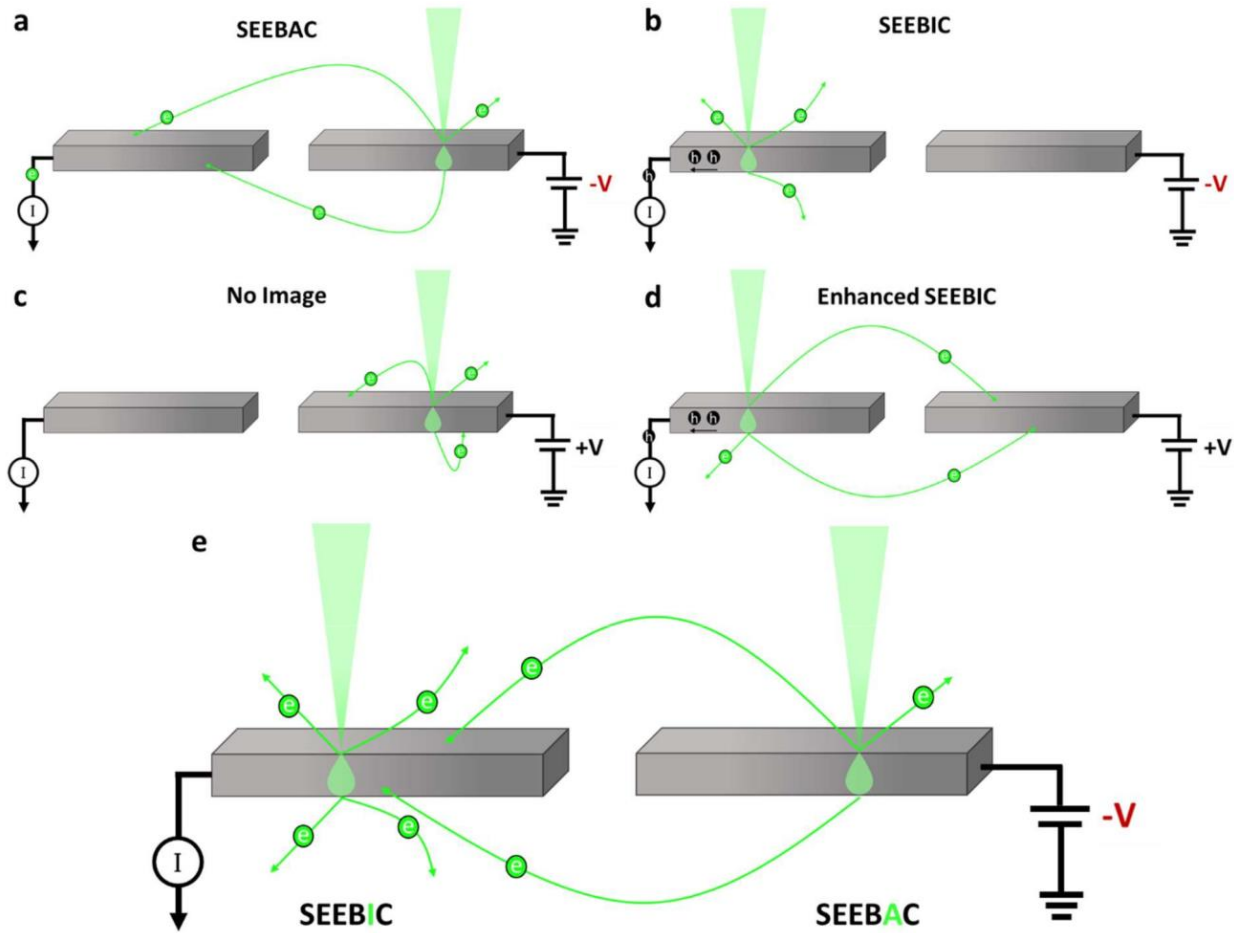

**Figure 11** Schematic demonstrating the different forms of imaging possible with two disconnected contacts and a voltage source. a) The e-beam is incident on a negatively biased specimen and the TIA is connected to an adjacent but electrically disconnected collector. SEs can be absorbed and an image can be formed. b) The e-beam is incident on the sensing/TIA electrode and the opposite/non-sensing portion of the specimen is negatively biased. Normal SEEBIC imaging occurs. c) The e-beam is incident on a positively charged non-sensing (disconnected from the TIA) portion of the specimen. Most SEs are recaptured and no image is formed. d) Enhanced SEEBIC resulting from the positive bias, reducing recapture of emitted SEs. e) SEEBIC and SEEBAC occurring under the same sample conditions as a result of e-beam position.

To experimentally demonstrate SEEBAC, devices were fabricated as previously and then Joule heated to clean the devices, cause $SiN_x$ substrate evaporation, and finally separation of the graphene ribbon at the constriction. A computer controllable voltage source was utilized to apply a bias to one half of the device.

Results are shown in Figure 12, where images were acquired with a bias of -10, 0, and +10 V applied to the electrode on the left hand side (LHS) of the device while the TIA was connected to the right hand side (RHS). In the region between the two sides, the SiN$_x$ substrate has been evaporated providing a vacuum intensity reference. When the specimen was negatively biased, Figure 12(a), a lower signal was observed than vacuum on the LHS and a decreased overall intensity was observed on the RHS. Notably, the lower-than-vacuum signal only occurred at the edges of the graphene where the field between the two halves is the greatest, enhanced SE emission is facilitated by lowering the work function, and capture on the opposing electrode is permitted due to proximity. Since electrons are flowing into the TIA instead of out, the contrast is reversed. When the specimen was positively biased, Figure 12(c), no contrast difference was observed between the LHS and vacuum but a pronounced increase in intensity was observed on the RHS.

Figure 12(d) clearly illustrates both SEEBAC and enhanced SEEBIC through a comparison of line profiles taken from the locations indicated by the dotted boxes in (a-c). When the LHS side was biased negatively, SEEBAC occurred at the edge of the biased graphene, which is clear since the signal is lower than that of vacuum. When the polarity was switched, no difference was observed between the biased side and vacuum, but an enhanced signal from the RHS electrode was observed when compared to zero bias, suggesting that recapture is suppressed and the best possible SEEBIC imaging conditions have been achieved.

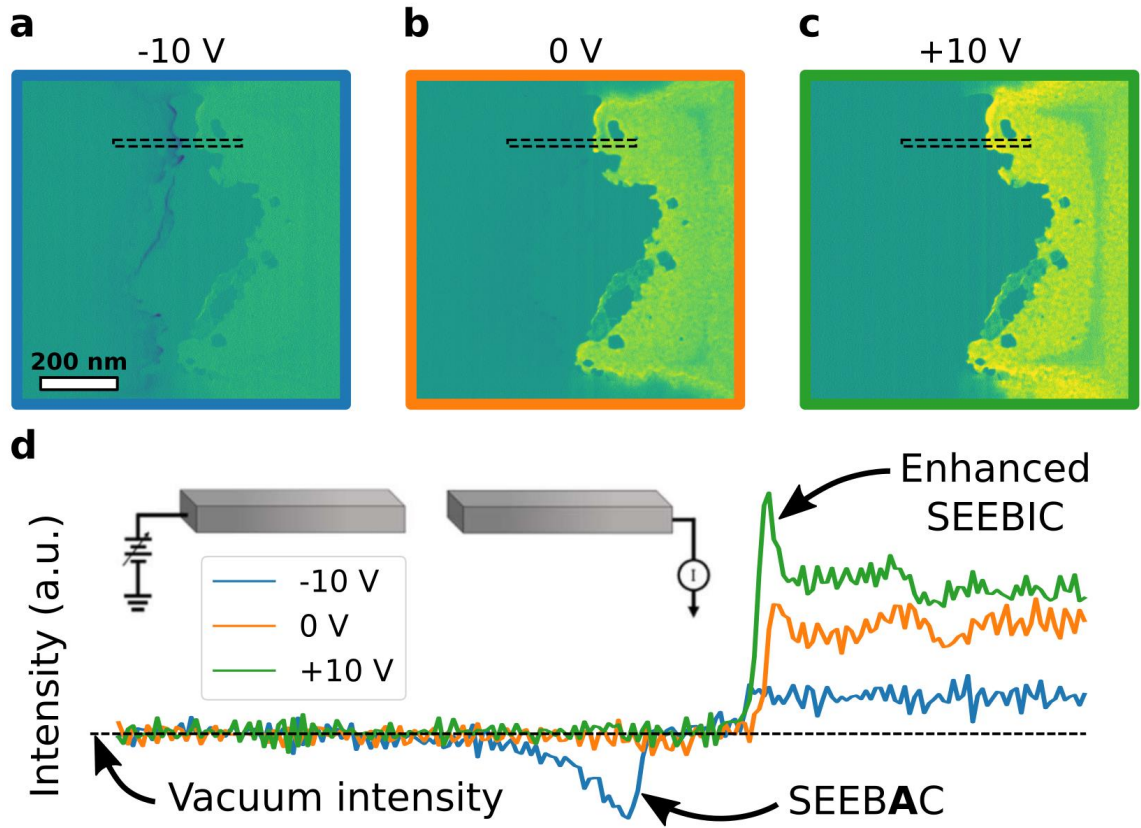

**Figure 12** a)-c) False color SEEBIC images acquired with bias voltages of -10, 0, and +10 V applied to the electrode on the left hand side (LHS) and the TIA connected to the right hand side (RHS). The displayed intensity range is the same for all three images to facilitate direct comparison. d) Integrated line profiles from SEEBIC images a)-c) taken from the areas shown by dashed rectangles. Schematic with the electrical connections overlaid. All three SEEBIC images were acquired at 200 kV and approximately 600 pA beam current and 300 µs dwell time.

**Resistive Contrast Imaging**

Resistive contrast imaging (RCI) can be used in EBIC configurations in both an SEM and STEM to elucidate the relative conductivity of specimen components (Smith et al., 1986). The contrast formation mechanism is straightforward and summarized schematically in Figure 13(a). Lower resistance regions appear brighter and higher resistance regions appear darker. RCI is most commonly used in SEM imaging as a failure analysis technique (Cole, 1989; Cho et al., 2009), but there have been recent reports of its use in STEM (Hubbard et al., 2020, 2019) as well as resistive switching imaging (Kamaladasa et al., 2015; Dyck, Swett, Evangeli, et al., 2021).

One reason STEM-based RCI has been infrequently reported is that the most common EBIC RCI mechanism is EBAC, which relies on absorption of the primary beam electrons. This is essentially a physical impossibility in the STEM, due to the high accelerating voltages and ultra-thin, electron-transparent specimens used; however, RCI can also be performed using SEEBIC instead of EBAC (Hubbard et al., 2019). Both EBIC, EBAC, and SEEBIC all generate a signal, which when connected via external circuits, can permit RCI. Knowledge of the type of induced current as well as the circuit configuration (e.g., grounded on both sides or biased on one) are necessary to properly interpret the images produced. One must also be cautious in interpreting the images since a change in contrast may be due to a discrete change in series resistance or a change in the entire circuit. Figure 13(a) illustrates this case, where the entire RHS of the specimen has a lower contrast despite the high resistance region being confined to a specific point. From an RCI perspective the SEEBIC analysis of graphene can be re-examined. It is well known that graphene grain boundaries represent high-resistance regions in an otherwise low-resistance material (Clark et al., 2013). The presence of a grain boundary could theoretically be detected with RCI in the STEM using SEEBIC as the charge generation mechanism, as demonstrated by the two examples of graphene devices exhibiting an abrupt change in SEEBIC contrast shown in Figure 13.

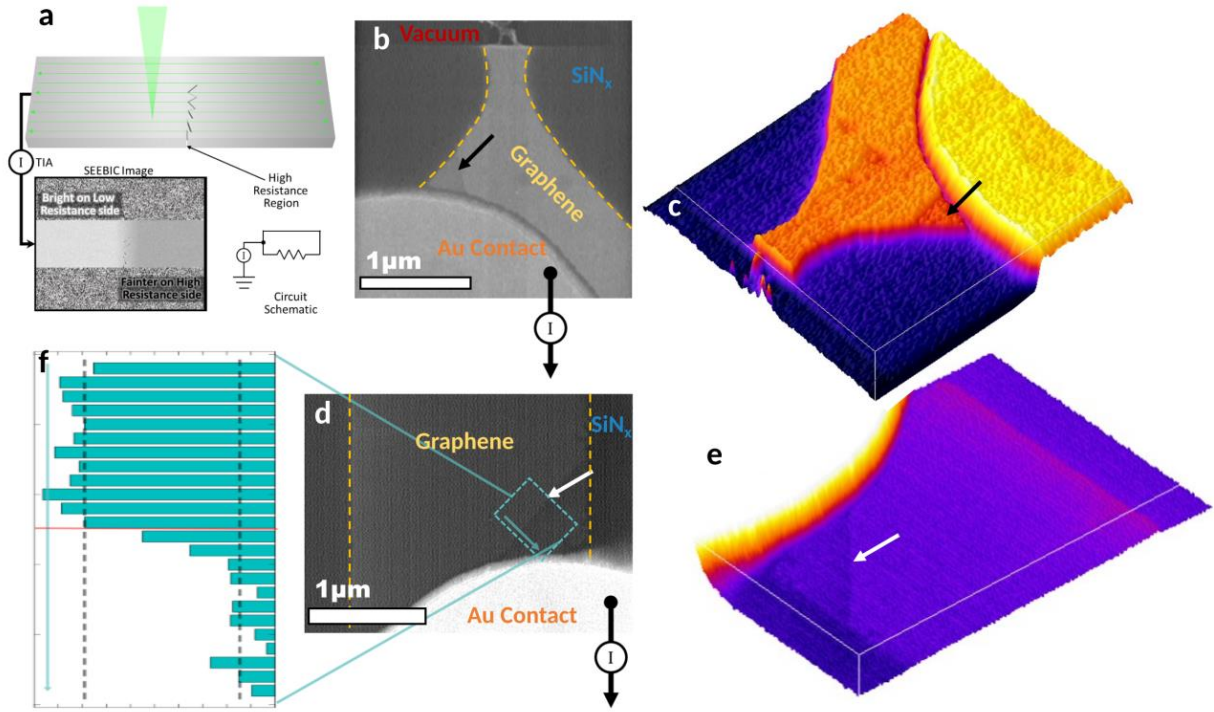

**Figure 13** Evidence for graphene grain boundary identified by RCI. a) Schematic illustrating the concept of RCI, where regions with lower resistivity and higher resistivity paths to ground are brighter and darker, respectively. b) SEEBIC image of a device where a portion of the graphene is clearly darker, c) 3D surface plot showing the same region indicated by the black arrow. d) Different device showing the same behavior, e) 3D surface plot of the same device with arrow indicating boundary, f) integrated line profile illustrating the drop in intensity across the boundary. Image in b,c) was acquired at 60 kV and 60 pA beam current and d,e) at 60 kV and 20 pA beam current.

Unfortunately, the regions of graphene were supported on the $SiN_x$ substrate so it was not possible to use the atomic resolution imaging capability of the STEM to probe whether a grain boundary was present at the edge of the contrast transition. Given the transfer process of the graphene onto the substrate and the abrupt height change from the metal contact to the $SiN_x$ membrane, it is also possible that the observed feature is a crack induced during the transfer process. Nevertheless, the imaging is suggestive of a change in resistance and further experiments could establish this as a technique for 2D materials and other nanostructures and could prove particularly useful for emerging techniques to sculpt and probe nanoelectronic devices in situ (Rodríguez-Manzo et al., 2016).

## Conclusions

In this work, we provided a brief overview of the current state of the field of STEM-based SEEBIC imaging. We have shown how STEM SEEBIC may be leveraged for characterizing 2D graphene-based nanodevice structures and provided detailed observations and discussions regarding the contrast generating mechanisms at work. Specifically, we illustrated how the significant contribution of the dielectric substrate plays a prominent role in visualizing supported graphene, demonstrated how SEEBIC can be used to reveal electrical discontinuities with high spatial resolution, showed that negative contrast is generated by a reversal in the current flow with bias, and presented evidence for the detection of graphene grain boundaries or partial electrical discontinuities through the use of SEEBIC-based RCI. These observations and discussions will serve to establish a useful foundation for the STEM SEEBIC technique and offer a productive pathway for future experiments.

## Acknowledgment

This work was supported by the U.S. Department of Energy, Office of Science, Basic Energy Sciences, Materials Sciences and Engineering Division (O.D. A.R.L., S.J.), and was performed at the Center for Nanophase Materials Sciences (CNMS), a U.S. Department of Energy, Office of Science User Facility (O.D.). J.A.M. was supported through the UKRI Future Leaders Fellowship, Grant No. MR/S032541/1, with in-kind support from the Royal Academy of Engineering. The authors acknowledge use of characterization facilities within the David Cockayne Centre for Electron Microscopy, Department of Materials, University of Oxford, alongside financial support provided by the Henry Royce Institute (Grant ref EP/R010145/1).

Competing interests: The authors declare none

ARGANDA-CARRERAS, I., KAYNIG, V., RUEDEN, C., ELICEIRI, K. W., SCHINDELIN, J., CARDONA, A. & SEBASTIAN SEUNG, H. (2017). Trainable Weka Segmentation: a machine learning tool for microscopy pixel classification. *Bioinformatics* **33**, 2424–2426.

BAILEY, R. S. & KAPOOR, V. J. (1982). Variation in the stoichiometry of thin silicon nitride insulating films on silicon and its correlation with memory traps. *Journal of Vacuum Science and Technology* **20**, 484–487.

BLÁZQUEZ, O., LÓPEZ-VIDRIER, J., HERNÁNDEZ, S., MONTSERRAT, J. & GARRIDO, B. (2014). Electro-optical Properties of Non-stoichiometric Silicon Nitride Films for Photovoltaic Applications. *Energy Procedia* **44**, 145–150.

BLELOCH, A. L., HOWIE, A. & MILNE, R. H. (1989). High resolution secondary electron imaging and spectroscopy. *Ultramicroscopy* **31**, 99–110.

BROWN, H. G., D'ALFONSO, A. J. & ALLEN, L. J. (2013). Secondary electron imaging at atomic resolution using a focused coherent electron probe. *Physical Review B* **87**, 054102.

BROWN, P. D. & HUMPHREYS, C. J. (1996). Scanning transmission electron beam induced conductivity investigation of a Si/Si1−xGex/Si heterostructure. *Journal of Applied Physics* **80**, 2527–2529.

CHO, S. J., KIM, T. E., HONG, J. K., HONG, J. T., KIM, H. S., HAN, Y. W., KWON, S. D. & OH, Y. S. (2009). Logic failure analysis 65/45nm device using RCI amp; nano scale probe. In *2009 16th IEEE International Symposium on the Physical and Failure Analysis of Integrated Circuits*, pp. 50–53.

CLARK, K. W., ZHANG, X.-G., VLASSIOUK, I. V., HE, G., FEENSTRA, R. M. & LI, A.-P. (2013). Spatially Resolved Mapping of Electrical Conductivity across Individual Domain (Grain) Boundaries in Graphene. *ACS Nano* **7**, 7956–7966.

COLE, E. I. (1989). Resistive contrast imaging applied to multilevel interconnection failure analysis. In *Proceedings., Sixth International IEEE VLSI Multilevel Interconnection Conference*, pp. 176–182.

CRETU, O., LIN, Y.-C. & SUENAGA, K. (2015). Secondary electron imaging of monolayer materials inside a transmission electron microscope. *Applied Physics Letters* **107**, 063105.
36